\documentstyle[aps,floats,graphicx]{revtex}
\begin{document}

\newcommand{\bec}{\begin{center}}
\newcommand{\ec}{\end{center}}
\newcommand{\be}{\begin{equation}}
\newcommand{\ee}{\end{equation}}
\newcommand{\beqn}{\begin{eqnarray}}
\newcommand{\eeqn}{\end{eqnarray}}
\newcommand{\bet}{\begin{table}}
\newcommand{\ent}{\end{table}}
\newcommand{\bib}{\bibitem}

\wideabs{

\title{
Does the thermal-spike affect low-energy ion-induced interfacial mixing?
}

\author{P. S\"ule, M. Menyh\'ard, K. Nordlund} 
  \address{Research Institute for Technical Physics and Material Science,\\
Konkoly Thege u. 29-33, Budapest, Hungary, sule@mfa.kfki.hu\\
Accelerator Lab., Helsinki, Finland
}

\date{\today}

\maketitle

\begin{abstract}
Molecular dynamics simulations have been used to obtain the three-dimensional
distribution of interfacial mixing and cascade defects in Ti/Pt multilayer system
due to single $1$ keV $Ar^+$ impact at grazing angle of incidence.
The Ti/Pt system was chosen because of its relatively high heat of mixing in the binary
alloy and
therefore a suitable candidate for testing the effect of heat of mixing on ion-beam mixing.
However, the calculated mixing profile is not sensitive to the heat of mixing. Therefore
the thermal spike model of mixing is not fully supported under these irradiation conditions.
Instead we found that the majority of mixing occurs after the thermal spike during the
relaxation process.
These conclusions are supported by liquid, vacancy as well as adatom analysis. 
The interfacial mixing is in various aspects anomalous in this system:
the time evolution of mixing is leading to a phase delay for Ti mixing,
and Pt exhibits an unexpected double peaked mixing evolution. 
The reasons to these effects are discussed.
\end{abstract}
}


\section{Introduction}

With the appearance of nanotechnology the importance of thin films
increased considerably. Either during making thin films by any kind of
various sputtering processes or in thin film analysis (various kinds of
sputter depth profiling techniques) the thin film system is subjected to
low energy ion bombardment. For this reason the study of the effects
of the low energy ion irradiation of thin film structures is of high
importance. 

 One of the effects of ion bombardment is interfacial ion beam mixing. This term
is used for that process when an originally sharp interface (IF) between pure A
and B materials gets broader due to ion bombardment. The ion mixing (IM) 
phenomenon is known for a long time \cite{AverbackRubia}. The majority of our knowledge refers to
cases when the interface to be mixed is relatively deep in the matrix (that
is, far from the free surface) and consequently the ion energy applied is in
the range of 100 keV \cite{Cheng}.

 Several theoretical approaches and
techniques have also been developed and a rather coherent picture seems to
have emerged \cite{AverbackRubia,Cheng}.
Though there is no principal difference between the low and high energy ion
mixing still the previous experimental and theoretical techniques cannot be
directly applied for the low energy case. The most important difference between the high and low
energy ion mixing is that (i) the latter occurs close to the free surface and
(ii) its extent is obviously much less than that of high energy mixing.

 With the increasing computation speed MD simulations have become a standard tool to study
the low energy ion beam surface interaction, and interface mixing as well.
MD simulation of IM has been the subject of several studies in the last decades \cite{AverbackRubia}.
IM in multilayer system has also been investigated by several groups 
using MD \cite{AverbackRubia,Urbassek,Nordlund,Kornich}.
A common feature of these studies is that only a single ion impact has been considered
during the simulations.
Experimental analysis of interfacial mixing in multilayer metals, however, always deal with
a series of ion impacts.
For instance, Auger depth profiling analysis of multilayers should lead to the removal of the target
during analysis \cite{Menyhard1}. 
As a first step, we in this study focus on understanding the single-ion process.
To go beyond this level we will employ a series of ion impacts in the next study.

 The other common feature of the studies available is that they use a normal angle
of incidence.
We should like to model ion milling and destructive depth profiling techniques \cite{Menyhard1}. Based on
experimental evidences we know that for the optimum application of
these methods one should apply low energy ion bombardment with grazing
angle of incidence (to minimize the surface morphology development). This
arrangement, however, is far from the previous one. First, the angle of
incidence is grazing in contrast to normal angle of incidence. 
Second, Pinzon {\em et al.} \cite{Pinzon} recently
showed by MD calculation that the damage structure changes with angle of incidence. At grazing angle of incidence
ion bombardment the defects are confined close to the surface.  

  Another interesting aspect of ion-beam mixing is theoretical.
It is still unclear whether the mechanism of IM is purely ballistic or heat spikes
play essential role in the magnitude of interfacial mixing \cite{AverbackRubia,Cheng}.
In particular within the thermal spike (TS) model \cite{Cheng} the heat of mixing ($\Delta H_m$) is 
found to be a key quantity in determining the magnitude of mixing.
The role of heat of mixing on governing mixing processes is questioned by
R. Kelly {\em et al.} \cite{Kelly}.
They emphasized the role of residual defects in mixing (long ranged effects).
Although the thermal spike concept is widely accepted, its applicability
for low-energy bombardments is not straightforward.
MD simulation of cascades have been reported with evidence that local
melting occurs during the termal spike phases which persists for
several ps in the $3-5$ keV energy regime \cite{MD_spike}.
Our intention is to study IM in the case of low energy impact at $1$
keV. This energy regime is around the lower limit of the TS model \cite{Averback},
and the number of simulation studies in this  field is limited
\cite{UrbassekPRB95}. Gades and Urbassek \cite{UrbassekPRB95} reported MD simulations for pure Cu at 1 keV which supports the phenomenological model of Cheng {\em et al.}. However, it is not clear whether the TS model will apply for
multilayers with different constituents at low ion energies. The reason we are particularly interested in 1 keV Ar 
bombardment is its relevance to Auger depth profiling \cite{Menyhard1}. 

   The reported MD simulations, in accordance with the model of Cheng {\em et al.} \cite{Cheng},
show that the interfacial mixing (IFM) exhibits an inverse-square dependence on the cohesive energy; however
they also show a non-linear variation of the interface broadening with $\Delta H_m$.
It is therefore interesting to study the effect of thermodynamic
properties such as the heat of mixing $\Delta H_m$ on mixing at low ion energies.
Studying high energy IM of Ti/Pt Kim {\em et al.} \cite{Kim} concluded that the mixing is governed by TS. Cirlin {\em et al.} \cite{Cirlin} measured the depth resolution of AES depth profiling on Ti/Pt system and concluded that even at 1 keV the mixing is partially determined by TS.

 In this communication we report on MD calculation of Ti/Pt layered structure
applying grazing angle of incidence $Ar^+$ bombardment at 1 keV. Though the crystal
lattice and the mass of atoms will be determined according to the basic feature of Ti
and Pt, the Pt-Ti mixed part of the potential will be varied to determine
the effect of $\Delta H_m$ on the mixing.

\section{The multilayer sample and the simulation method}

Classical molecular dynamics simulations were used to simulate collisional
mixing using the PARCAS code developed by Nordlund {\em et al.} \cite{Nordlund_ref}.
Here we only shortly summarize the most important aspects.
Its detailed description is given in  \cite{Nordlund,Nordlund_ref}.
The classical equations of motion were solved according to the
Gear fifth-order predictor-corrector algorithm.
A velocity dependent variable time step of integration was used.
Periodic boundary conditions were applied to the fixed-size
simulation cells.
The electronic stopping power was included in the runs as a nonlocal frictional
force affecting all atoms with a kinetic energy higher then 10 eV.

  The sample consists of 37289 atoms for the interface system
with 4 Ti top layers and a bulk which is Pt.
The Ti (hcp) and Pt (fcc) layers at the interface are separated by $2.8$ \AA.
The computational cell has a size of roughly $85 \times 85 \times 85$ \AA.
This simulation cell is of our particular interest because preliminary
calculations indicate that 1 keV Ar$^+$ with grazing angle of incidence
provides the interesting situation when the recoils penetrates approximately
4 layers deep below the surface. Therefore the average center of the cascade is positioned close the interface.
 The lattice mismatch is optimized properly at the interface in order to reduce the
interfacial strain to less then $2$ \%. The best match of Ti (hcp) and
Pt (fcc) layers are carried out by probing various structures put together by hand.
The most favorably matched system have been selected for further simulations.

 The crystallite is kept at 0 K at the beginning of the simulations.
The entire iterfacial system is equilibrated prior to the irradiation
simulations and the temperature scaled softly down towards zero at the
outermost three atomic layers during the cascade events \cite{Nordlund_ref}.
Since in the present study we are interested in the fundamental
mechanism of interfacial mixing, we choose 0 K as the ambient temperature seems
to be appropriate to prevent any thermally activated effects from complicating the
analysis. A study of temperature dependence of IFM is planned in a further study.
The collision cascade were initiated by placing an Ar atom
outside the surface and giving it a kinetic energy of $1$ keV towards
the surface. The initial velocity direction of the
impacting atom was $83$ degree with respest to the surface normal.

  To obtain a representative {\em statistics}, 
the impact position of the incoming ion is varied
randomly up to $15$ events.
We find considerable variation of mixing as a function of the impact position of the recoil. 
The simulations are terminated at about $20$ ps after the projectile impact.

  During the simulations, an atom was labeled "liquid" if the average kinetic
energy of it and its nearest neighbours was above an energy corresponding to the
melting point of the material through the usual relation between temperature
and kinetic energy, $E=\frac{3}{2} kT$.
It is then possible to carry out liquid atom analysis, which allows us to
estimate the spatial extent of the cascade or the thermal spike region (the local melt).
The liquid phase (typically occurs at $0.1-0.5$ ps with a sharp "ballistic" peak \cite{AverbackRubia,Nordlund})
is the direct result of thermalization of ballistic recoils and in this sense
these liquid atoms do not correspond to the classical concept of a liquid in any
meaningful way \cite{NordlundPRB98}.
However, if the temperature of the atoms persists above the melting temperature
longer then 1-2 ps then it can be called as a liquid which corresponds to a thermal spike.
The liquid region corresponds to a classical liquid in the sense that the
pair distribution function resembles that of a liquid in equilibrium
\cite{NordlundPRB98}.

 The Berendsen temperature control is used throughout the simulations \cite{Allen}.
A complete analysis of the structural output file (movie file) has been carried out for 
each irradiation event for some chosen time steps (mixing, liquid, vacancy, adatom etc. analysis).

 An atom is labeled mixed if it moves beyond the interface during the simulation by more then the
half of the Ti-Pt interfacial distance in the initial sample ($\sim 1.4 \AA$).

 Vacancies were recognized in the simulations using a very simple analysis.
A lattice site with an empty cell was considered to be a vacancy where the diameter of the cell sphere
is set to $\sim 2 \AA$ (the radius of a sphere around the relocated atom around its original position) which is about $60 \%$ of the average
atomic distance in this system.
We find this criterion is suitable for counting the number of vacancies.
We monitored
the change in the surface location, and it is less than $0.3 \AA$ during the simulation time.
It is therefore does not affect the efficiency of the method used for vacancy, adatom
and mixing analysis.


\begin{figure}[hbtp]
\begin{center}
\includegraphics*[height=6cm,width=7cm,angle=0.0]{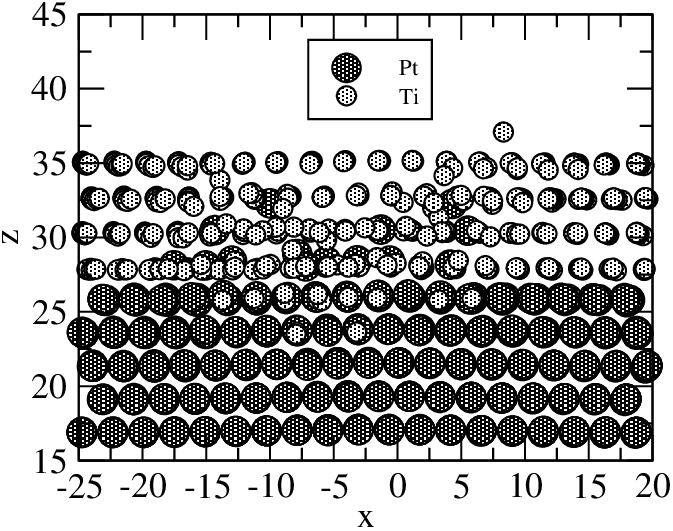}   
\caption[]{
Cross-sectional view at the Ti/Pt interface system at the end of the simulation
($\sim 20$ ps).
The coordinates are in $\AA$ unit.
(The upper 4 layers were originally Ti and the bulk is Pt.)
The cross section is taken along the $z$-axis (depth profile).
Heavily mixed sample taken from statistics ($N_{mix} \approx 52$).
}
\label{pict}
\end{center}
\end{figure}


{\hspace{-3cm}

\begin{figure}[hbtp]
\begin{center}
\includegraphics*[height=6cm,width=7cm,angle=0.0]{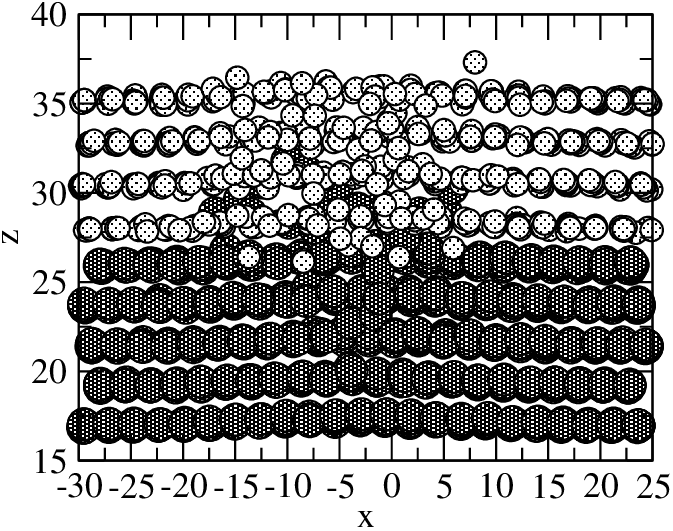}
\caption[]{
Cross-sectional view at the Ti/Pt interface system at the begining of the cooling
period
($\sim 2$ ps).
Heavily mixed sample taken from statistics ($N_{mix} \approx 52$).
}
\label{pict2}
\end{center}
\end{figure}
}



\begin{figure}[hbtp]
\begin{center}
\includegraphics*[height=6cm,width=7cm,angle=0.0]{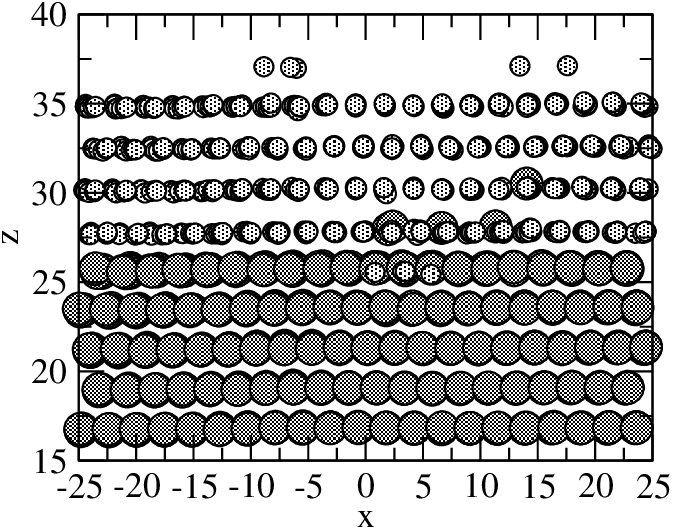}   
\caption[]{
Cross-sectional view at the Ti/Pt interface system at the end of the simulation
($\sim 20$ ps) for the low mixing sample ($N_{mix} \approx 9$).
}
\label{pict3}
\end{center}
\end{figure}


{\hspace{-3cm}


\begin{figure}[hbtp]
\begin{center}
\includegraphics*[height=6cm,width=7cm,angle=0.0]{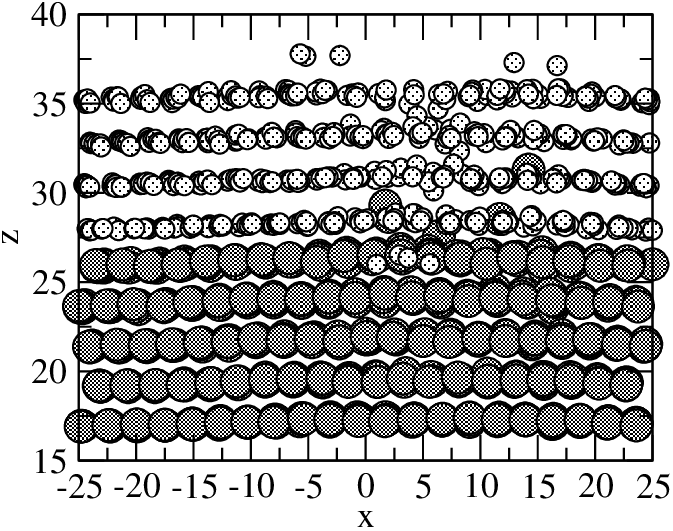}
\caption[]{
Cross-sectional view at the Ti/Pt interface system at the begining of the cooling
period
($\sim 2$ ps).
Weakly mixed sample taken from statistics ($N_{mix} \approx 9$).
}
\label{pict4}
\end{center}
\end{figure}
}


\newpage


\begin{figure}[hbtp]
\begin{center}
\includegraphics*[height=6cm,width=7.0cm,angle=0.0]{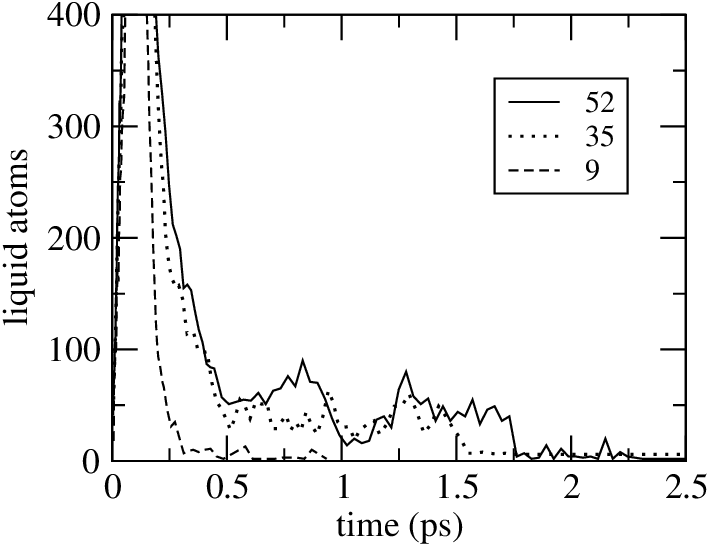}
\caption[]{
The time evolution of the thermal spike.
The number of liquid atoms as a function of time under 1 keV Ar+ bombardement
for low ($N_{mix} \approx 9$) and high degrees of interface mixing ($N_{mix} \approx 35,52$; $\Delta H_m
\approx 75$ kJ/mol).
}
\label{fig_liq}
\end{center}
\end{figure}


 Our simulation method \cite{Nordlund_ref} used in this study allows us
to assign the value of heat of mixing of the metal pair relatively easily
.
 The attractive part of 
the Cleri-Rosato many-body potential \cite{CR} used in this study
\be
V^i(r_{ij})=-\biggm(\sum_j \xi^2 e^{-2 Q (r_{ij}/r_0-1)}\biggm)^{1/2},
\label{xi}
\ee
where $r_{ij}$ represents the interatomic distance between atoms $i$ and $j$, 
$r_0$ is the first-neighbour distance.
The total cohesive energy of the system is then
\be
E_c=\sum_i(E_R^i+E_B^i),
\ee
where $E_R^i$ is a repulsive term of the Born-Mayer type \cite{CR,Eckstein}.


\begin{figure}[hbtp]
\begin{center}
\includegraphics*[height=6cm,width=7.0cm,angle=0.0]{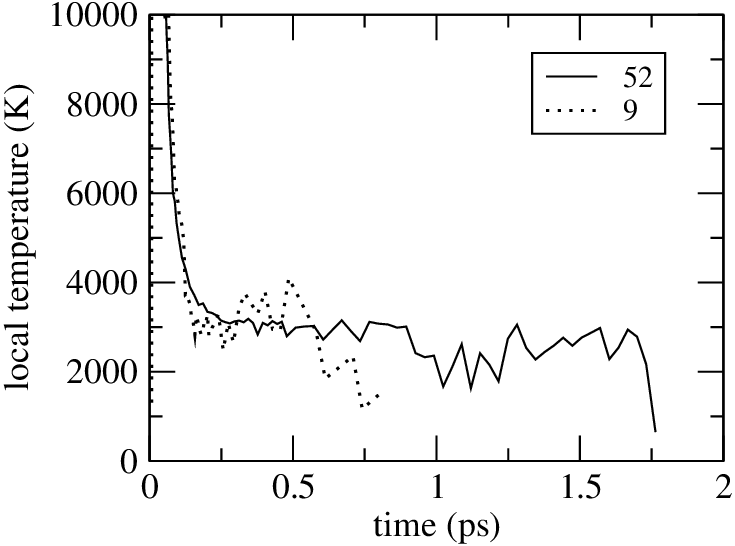}
\caption[]{
the time evolution of the local temperature (K) in the thermal spike
for the low and high degrees of interface mixing ($N_{mix} \approx 9$ and $52$).
}
\label{temp_time}
\end{center}
\end{figure}


  By varying the parameter $\xi$ in Eq.~(\ref{xi}) the heat of mixing can easily be varied
without influencing other parameters of the crystall. 
The heat of mixing is calculated in the usual way,
\be
\Delta H_m = 0.5 (E_A+E_B)-E_{AB},
\label{HM}
\ee
where $E_A$, $E_B$ and $E_{AB}$ are the total energies/atom of the corresponding
constituents.
The heat of mixing of Ti/Pt is substantial ($-75$ kJ/mol) according to the Miedema calculation \cite{cohesion,Miedema}.
Varying $\xi$ in the range of $0-2.55$ the heat of mixing calculated from Eq.~(\ref{HM})
varies in the range of $+75 - 141$ kJ/mol, the Miedema value can be achieved by taking $\xi \approx 2.35$.

\section{Single ion impact on the Ti/Pt system}

   The most important technical problem of the simulation is that the various parameters
describing the change of the sample due to ion impact including mixing strongly depends
on the position of the ion impact. Thus we always consider $15$ events with random impact
positions and the average and the scatter of the parameters will be given.
FIGs~(\ref{pict})-(\ref{pict4}) show final and intermediate states of the samples after 
$20$ and at $2$ ps after a single ion impact for $\xi=2.35$. 
FIGs~(\ref{pict})-(\ref{pict2}) depict
a strongly and (FIGs~(\ref{pict3})-(\ref{pict4})) a weakly mixed sample with different impact
positions.
Originally the upper 4 layers were pure Ti. For the present case we can conclude
that a single ion causes a heavy distortion in the sample. Pt and Ti appears in the originally
pure Ti and Pt regions, respectively, that is, considerable mixing occurs.
Besides the mixed atoms we can recognize several vacancies, interstitials and adatoms.
For the shown cases in these events the number of mixed atoms, $N_{mix} \approx 52$ and $9$, respectively.


\begin{figure}[hbtp]
\begin{center}
\includegraphics*[height=6cm,width=7cm,angle=0.0]{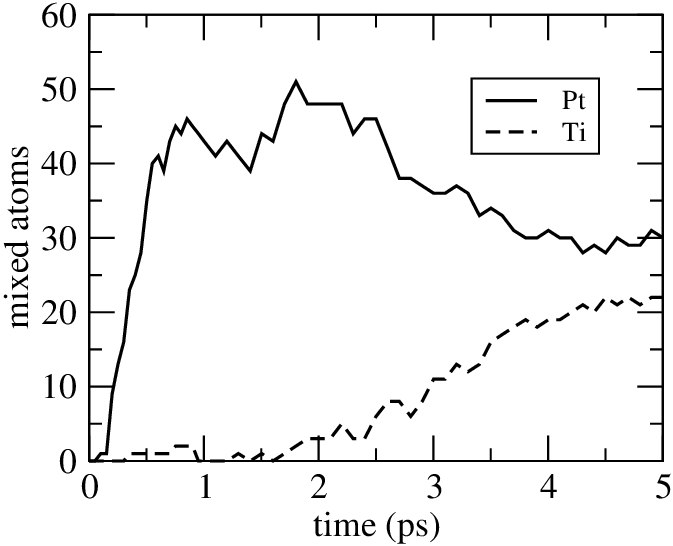}
\caption[]{
The evolution of mixing as a function of simulation time (ps)
for Pt and Ti at large mixing rate ($N_{mix} \approx 52$, $\Delta H_m \approx 75$ kJ/mol).
}
\label{mix_high}
\end{center}
\end{figure}


Considering all $15$ events $N_{mix}$ varies between $9-52$.
Comparing FIGs~(\ref{pict})-(\ref{pict2}) we see the heavily distorted structure at $2$ ps
and the considerable mass transport of Pt in the Ti phase.
Comparing FIGs~(\ref{pict3})-(\ref{pict4}) the difference between the relaxed ($20$ ps) and 
the distorted structures is even more remarkable.
In particular FIG~(\ref{pict}) shows us that the system remained distorted 
even after relaxation.
From the point of view of the mechanism, however, these figures are not satisfactorily informative.

  If we would like to correlate IFM with the TS, we should consider time evolution of the
number of liquid atoms and the average local temperature of the liquid atoms. This is shown in FIGs~(\ref{fig_liq})-(\ref{temp_time}) for various events, in case for
low, medium and high mixing rates.
For simplicity we display only two mixing cases on FIG~(\ref{temp_time}).
The curves are rather similar.
All are dominated by a huge peak at around $150$ fs (characteristic ballistic peak).
The thermal spike period can be attributed \cite{Rubia,Averback} to the tails of the curves.
Its length and magnitude depends on the number of mixed atoms. The higher the number of mixed
atoms the longer is the TS period. However, all liquid atoms disappear at around $1.5$ ps.
Thus if the mixing is governed by TS then it should occur until $\sim 1.5$ ps.
The average local temperature of the liquid atoms 
(FIG~(\ref{temp_time})) remains above the melting temperature during the TS which
indicates that there is a local melt already at 1 keV which persists until 1-1.5 ps.
The number of liquid atoms abruptly scales down to zero at $t \approx 1.75$ ps (the end
of the TS) due to the solidification of the local melt.

\begin{figure}[hbtp]
\begin{center}
\includegraphics*[height=6cm,width=7cm,angle=0.0]{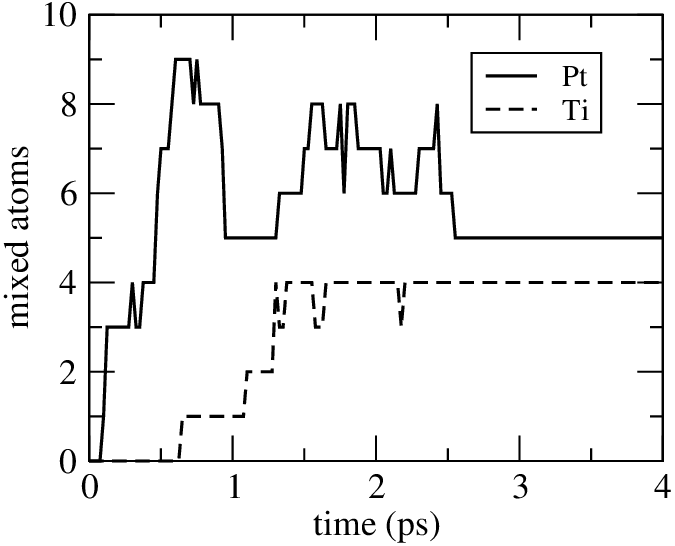}
\caption[]{
The evolution of mixing as a function of simulation time (ps)
for Pt and Ti at low mixing rate ($N_{mix} \approx 9$, $\Delta H_m \approx 75$ kJ/mol).
}
\label{mix_low}
\end{center}
\end{figure}

  In FIGs~(\ref{mix_high})-~(\ref{mix_low}) the time evolution of IFM, that is the number of mixed Ti and Pt atoms as a function of time is given for a weakly
and strongly mixed sample.
It is obvious that the two figures show similar features and thus we can conclude that the time
evolution does not depend on the magnitude of mixing.

These curves show some unexpected features. First of all mixing is strongly asymmetric.
In case of high mixing (FIG~(\ref{mix_high})) the number of the mixed (heavier) Pt atoms starts to increase just
after the ballistic process is terminated and reaches a maximum during the TS period ($\sim 1$ ps).
With the decrease of the number of liquid atoms to zero at $\sim 1.75$ ps the number of mixed Pt atoms
also decreases. But after this local minimum an additional increase comes. It reaches about the same
value at about $\sim 2$ ps which is at $1$ ps but at $2$ ps there is no liquid atom present.
This feature is shown in an even more striking way in FIG~(\ref{mix_low}).
In the case of low mixing the first peak of Pt appears at about $0.5$ ps, while the number of
liquid atoms disappears at $\sim 0.25$ ps (see FIG~(\ref{fig_liq}).
On the other hand the general behaviour of the curves in FIG~(\ref{mix_high})
and FIG~(\ref{mix_low}) are similar and it seems that the time evolution of the number of mixed atoms
is independent of the magnitude of mixing for Pt.
However, we see somewhat different behaviour for Ti. In FIG~(\ref{mix_low}) Ti reaches its maximum
more rapidly then in FIG~(\ref{mix_high}).
The number of the mixed Ti atoms starts to increase when the number of mixed Pt atoms
reached a maximum. With the increase of the number of mixed Ti atoms the decrease of the number of
mixed Pt atoms takes place, however, we see no evidence for such a process for the weakly mixed sample
on FIG~(\ref{mix_low}).
During these processes the number of liquid atoms is zero.
To sum up we see the decoupling of mixing of Pt and Ti during the TS and coupling sets in after a while
the TS.

\begin{figure}[hbtp]
\begin{center}
\includegraphics*[height=6cm,width=7cm,angle=0.0]{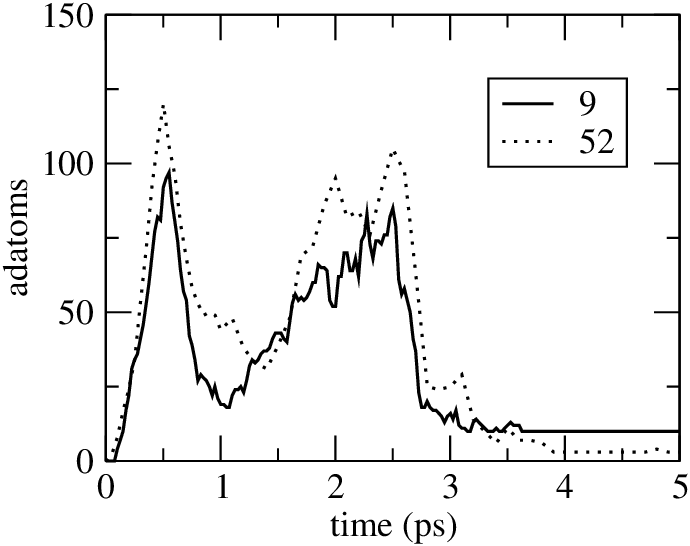}
\caption[]{
The evolution of the number of adatoms as a function of simulation time (ps)
at high and low mixing rates ($N_{mix} \approx 9, 52$).
}
\label{adatom}
\end{center}
\end{figure}

\begin{figure}[hbtp]
\begin{center}
\includegraphics*[height=6cm,width=7cm,angle=0.0]{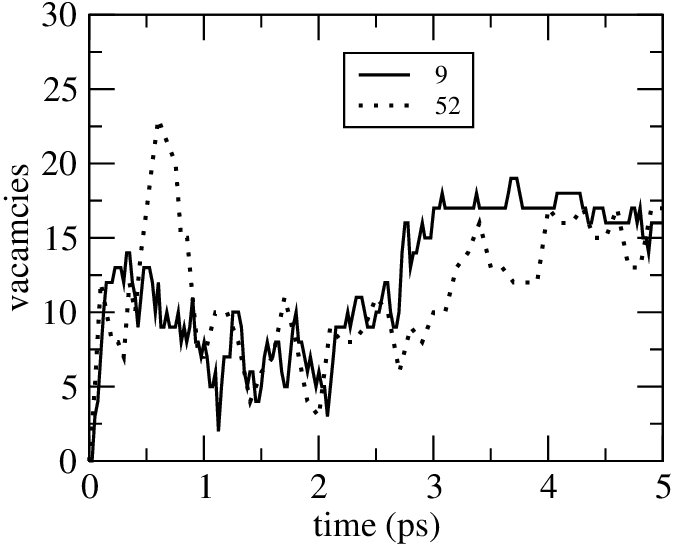}
\caption[]{
The evolution of the number of vacancies as a function of simulation time (ps)
at high and low mixing rates ($N_{mix} \approx 9, 52$).
}
\label{vacancy}
\end{center}
\end{figure}

 The time evolution of the adatoms, shown in FIG~(\ref{adatom}), somewhat resembles that of mixed Pt atoms. It is a double peaked curve, which abruptly decreases at $\sim 3$ ps when the number of mixed Pt atoms
reaches a quasi-constant value.
On the other hand the time evolution of the vacancies differs from that of adatoms (FIG~(\ref{vacancy})). 
There is a smaller peak in the vacancy time evolution in the $0-1$ ps range. Between $1-3$ ps the
number of vacancies reaches a minimum with a low value of $\sim 4$, and it begins to rise at $\sim3$ ps
to reach a constant value of about $\sim 20$ at $5$ ps.
We note that the increase of the number of mixed Ti atoms starts somewhat earlier (at $2$ ps)
and reaches a constant value at $3$ ps.


{\em Based on these observations it seems to be clear that a simple TS model cannot account for
the mixing process}. Rather it seems that the mixing occurs as a relaxation process.
Surprisingly, IFM starts with a considerable mass transport of Pt and partly coincides with the TS,
but as it is shown for the weakly mixed sample (FIG~(\ref{mix_low})), it might be independent from it.

The reasonable correlation with the number of the Ti adatoms suggests that the IFM process
of Pt is strongly influenced by the presence of the free surface.
After the first phase of Pt mixing is over the Ti atoms starts to move and with some delay considerable
number of vacancies appear.
Both changes can be explained as a relaxation process, which decreases the energy of the system.

\begin{figure}[hbtp]
\begin{center}
\includegraphics*[height=6cm,width=7cm,angle=0.0]{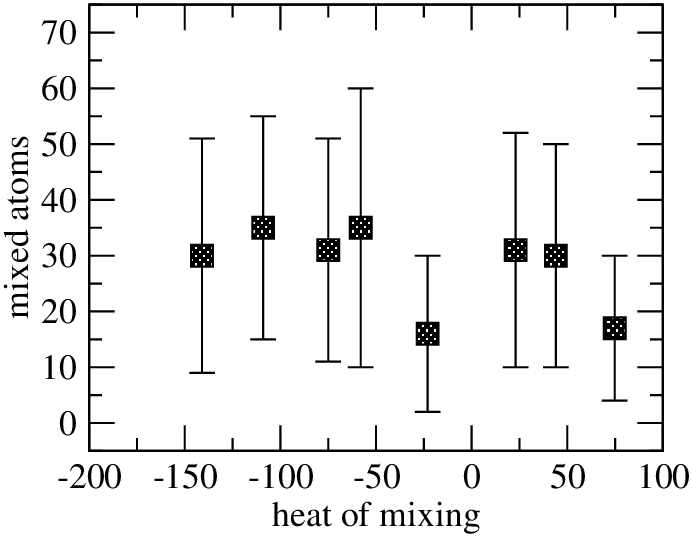}
\caption[]{
The total number of mixed atoms as a function of heat of mixing
in the first irradiation step as obtained varying $\xi$ from
$0.0$ to $2.55$. 
The error bars indicate the scatter obtained within $15$ events
at each point. The lower and the upper point of the error bar
correspond to the lowest and the highest number of mixed atoms.
}
\label{homm}
\end{center}
\end{figure}

\begin{figure}[hbtp]
\begin{center}
\includegraphics*[height=6cm,width=7cm,angle=0.0]{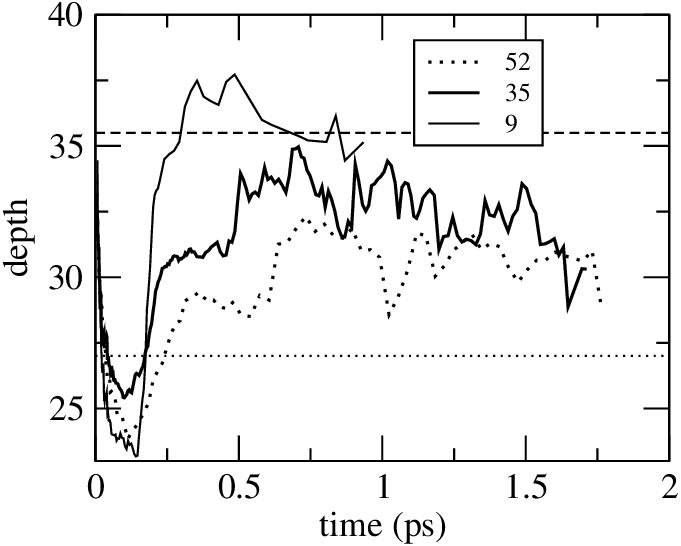}
\caption[]{
The $z$-axis position (depth in $\AA$) of the center of the liquid zone (thermal spike) as a function of the
simulation time for 3 different mixing rates obtained from liquid atom analysis
(only hot atoms are considered without their compact neighbourhood).
The dashed and dotted horizantal lines are the indication of the free surface and the interface.
}
\label{liq_z}
\end{center}
\end{figure}

  Based on the above discussion we might conclude that the IFM is independent of the TS. If it is so, then
the mixing cannot depend on the heat of mixing. With the change of the Ti-Pt interaction potential by means of
changing $\xi$ we can produce systems which are basically similar is mass, crystal structure etc.
but their heat of mixing is different.
FIG~(\ref{homm}) shows the dependence of the number of mixed atoms on $\Delta H_m$ based on 15 simulations
with random impact positions/point.
As we have already mentioned, the magnitude
of mixing is strongly impact position dependent.
Anyhow, having this statistics we cannot see any dependence on $\Delta H_m$, which supports the discussion
above although the high rate of incertainties at each of the points seen in FIG~(\ref{homm}) does not
prove solidly our findings above. 
In particular our findings are in contrast with the results of Gades {\em et al.}
\cite{UrbassekPRB95}. They found a strong dependence of mixing on $\Delta H_m$ in agreement
with the phenomenological models \cite{Cheng}.
Although that study is confined to the case of mixing of pure Cu and of various model metals
having the same crystall structure
and therefore the unsymmetrical nature of mixing does not occur.
What could be the role of TS if any in this mixing process?
To answer this question we also show the time evolution of the center of the liquid zone in
FIG~(\ref{liq_z}). The dotted lines at $z=35.5$ and at $27$ ($z$ is the depth position in $\AA$) mark the free surface and the interface,
respectively. For higher mixing rate the center of the liquid zone (the average of the $z$-coordinate
of the hot atoms) is nearly confined close to the interface, while for the case of lowest mixing it moves towards the surface.
As it was mentioned in the case of the weakly mixed sample, the magnitude of the overall mixing
scales with the duration of the liquid phase.
Since the center of the liquid phase is at the interface, one might argue that considerable energy
is concentrated around the interface which later partly after the freezing of the TS
relaxes causing IFM.


But the liquid phase energy depends only on the conditions of ion impact 
and will not change upon $\Delta H_m$. Thus we explained that though the liquid phase
creation is strongly connected to the magnitude of mixing, still the mixing is
independent of $\Delta H_m$.


\section{Conclusion}

 Molecular-dynamic simulation was carried out for the Ti/Pt system to explain the ion bombardment induced
mixing at 1 keV $Ar^+$ ion bombardment at grazing angle of incidence.
We have found that the mixing does not depend on the heat of mixing, as might have been expected.
That is, it is not the thermal spike process which governs the mixing in this case.
Though we found that the magnitude of mixing is scaling with the number and position of the liquid atoms,
this does not depend on the heat of mixing. Rather the majority of the mixing processes are taking place in that time,
when liquid atoms are not present. Moreover it was found that the mixing of Ti and Pt
are not correlated. We propose that the mixing occurs as a relaxation prcoess.

\section{acknowledgement}
{\small 
This work is supported by the OTKA grants F037710
and T30430
from the Hungarian Academy of Sciences and from EU contract no. ICAI-CT-2000-70029.}

\end{document}